\documentclass[12pt]{article}
\usepackage{etex}
\usepackage{epsfig}
\usepackage{epsf}
\usepackage{latexsym}
\usepackage{amsfonts,amssymb,amsmath} 
\usepackage{mathtools}
\usepackage{color}
\usepackage{cancel}
\usepackage[a4paper,vmargin=3cm,hmargin=2.1cm]{geometry}
\usepackage{multirow}
\epsfverbosetrue
\newcommand{\goth}[1]{\mathfrak{#1}}
\newcommand{\double}[1]{\mathbb{#1}}

\newcommand{\rr}{\double{R}}
\newcommand{\zz}{\double{Z}}

\newcommand{\gggg}{\goth{g}}

\newcommand{\dd}{\mathcal{D}}

\newcommand{\de}{\hbox{\rm{d}}}

\newcommand{\pa}{\partial}

\newcommand{\dpp}{\vcentcolon}
\newcommand{\bb}{\begin{eqnarray}}
\newcommand{\ee}{\end{eqnarray}}
\newcommand{\eee}{\nonumber\end{eqnarray}}
\newcommand{\qq}{\quad}

\begin{document}

\thispagestyle{empty}

\begin{center}
${}$
\vspace{2cm}

{\Large\textbf{The gauge theoretical underpinnings of general relativity}} \\

\vspace{1cm}

{\large
Thomas Sch\"ucker\footnote{
Aix Marseille Univ, Universit\'e de Toulon, CNRS, CPT, Marseille, France
\\\indent\qq
thomas.schucker@gmail.com\\\indent\qq
supported by the OCEVU Labex (ANR-11-LABX-0060) funded by the
"Investissements d'Avenir" 
\\\indent\qq
French government program
 }}

\vspace{2cm}

\hfill{\em To the memory of Christian Duval}

\vspace{2cm}

{\large\textbf{Abstract}}

\end{center}

The gauge theoretical formulation of general relativity is presented. We are only  concerned with local intrinsic geometry, i.e. our space-time is an open subset of a four-dimensional real vector space. Then the gauge group is the set of differentiable maps from this open subset into the general linear group or into the Lorentz group or into its spin cover.\\[2cm]
Contribution to the
Heraeus-Seminar ``100 Years of Gauge Theory''\\
organized by Silvia De Bianchi and Claus Kiefer\\
  30 July - 3 August 2018\\ 
  Physikzentrum Bad Honnef\\
  To appear in the proceedings ``100 Years of Gauge Theory. Past, present and future perspectives'' in the series `Fundamental Theories of Physics' (Springer).

\vspace{2cm}

\vskip 1truecm

\eject

\section{Introduction}

One of the many satisfying features of general relativity is -- at least for me -- that it allows for many different approaches, some of which are:
\begin{itemize}\item
field theoretic approach
\item
${\qq\ \, }$geo-metric approach
\item
chrono-metric approach
\item
gauge theoretic approaches
\item
causal approach
\item
perturbative approaches
\item
numerical approaches
\item
discretized approaches
\item
quantum approaches ...

\end{itemize}
An unpleasant corollary to the richness of having several approaches is that some colleagues get carried away by one of the approaches and ignore or dispraise other approaches. If  mountaineer $A$ in Figure 1 thinks that his approach to his favorite summit is unique, then maybe he is sitting on a pre-summit. And he may even be tempted to build a chapel there and to throw stones at his colleague $B$, when she tries to climb higher. 

\begin{figure}[h]
\begin{center}
\includegraphics[width=14.9cm, height=7.8cm]{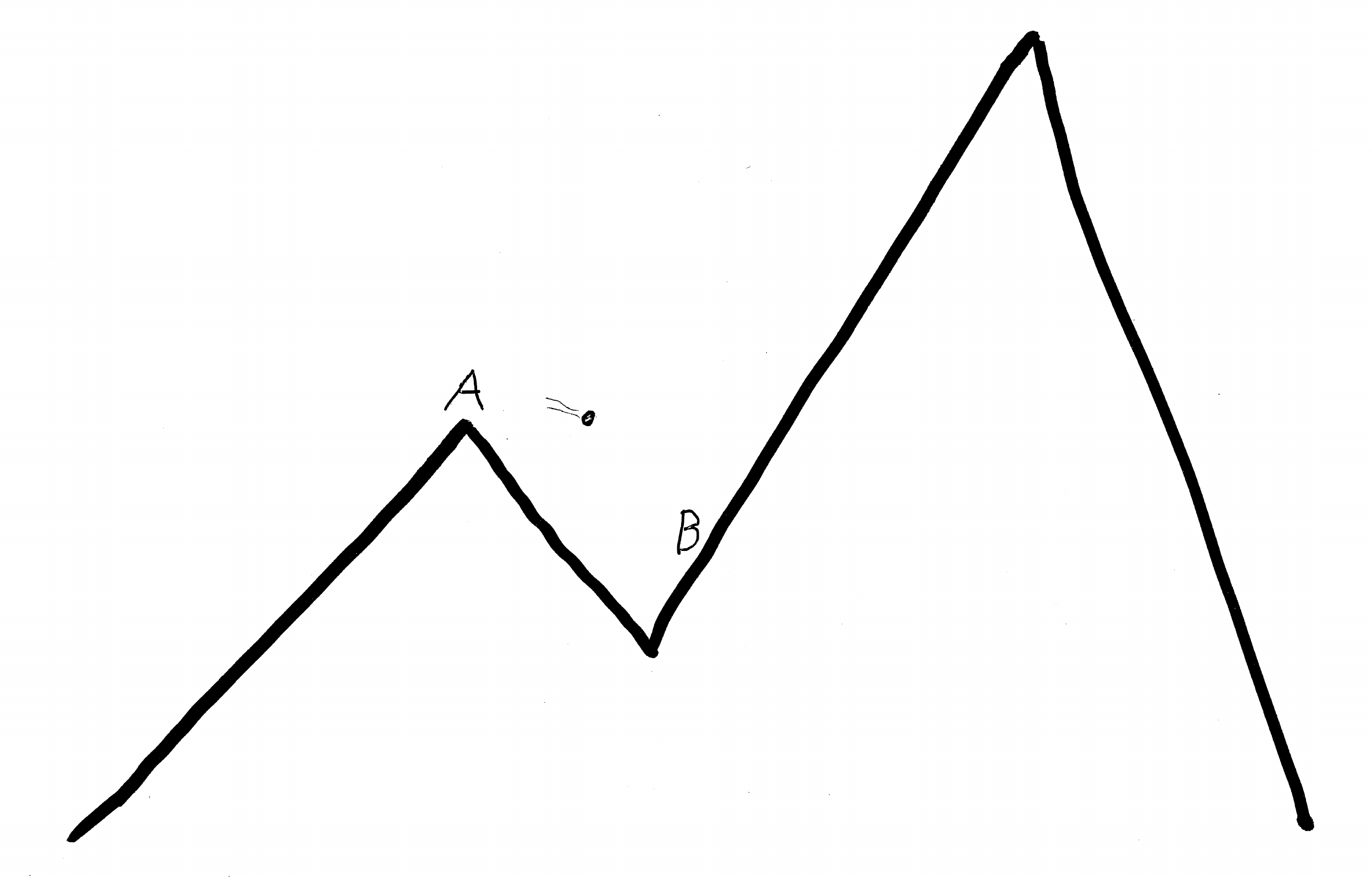}
\caption{Sitting on a pre-summit}
\end{center}
\end{figure}

The main subject of this contribution is a particular gauge theoretic approach to general relativity, another one being presented by Friedrich Hehl at this meeting. But let us start by saying a few words about the field theoretic, chronometric and geometric approaches.

\section{Field theoretic approach}

The field theoretic approach to general relativity is best set up in parallel to Maxwell's theory of electromagnetism.

Any electric charge $Q$ is source of an electric field, while mass,  or better energy $E$, is source of a gravitational field. The main difference is: charge is Lorentz invariant, while energy is only one component of  a 4-vector, the other three components being the momentum $\vec p$. Therefore Einstein's theory has systematically one index more than Maxwell's.
\bb 
Q&{\color{blue}\mbox{sources}}&
p^\mu =(E/c,\vec p) \nonumber\\[1mm]
 j^\nu=(c\rho,\vec{j})&{\color{blue}\mbox{densities}}&
 T_{\mu\nu}
\nonumber\\[1mm]
A^\mu=(V/c,\vec A)&{\color{blue}\mbox{fundam. fields}}&
{g_{\mu\nu}}
\nonumber\\[1mm]
\dd_2A=j&{\color{blue}\mbox{field eq.}}&
{\dd_2g=T}
\nonumber\\
\!\!\!\!\!\!\!\!\!\!\!\!\!\!\!\!\!\!\!\!\!\!(\dd_2A)^\mu\dpp =
\epsilon_0c^2\pa_\nu(\pa^\nu A^\mu
-\pa^\mu A^\nu)&\!\!\!{\color{blue}\mbox{diff. oper.}}&{
(\dd_2g)_{\mu\nu}\dpp =\,\frac{c^4}{8\pi G}\,
(R_{\mu\nu}-{\frac{1}{2}}\,R\,g_{\mu\nu}-\Lambda\, g_{\mu\nu})}
\nonumber\\
\dot {}\dpp =d/d\tau,\qq
m\ddot{x}^\mu-q{F^\mu}_\nu\dot{x}^\nu=0&
{\color{blue} \mbox{test part.}}&
{
{\cancel{m}}
\ddot{x}^\lambda+
{\cancel{m}}
\,{\Gamma^\lambda}_{\mu\nu}\dot x^\mu\dot x^\nu=0}
\nonumber\\[1mm]
F_{\mu\nu}\dpp =\pa_\mu A_\nu-
\pa_\nu A_\mu&{\color{blue}\mbox{auxiliaries}}&
{
{\Gamma^\cdot}_{\cdot\cdot}=(g^{-1}\pa)^\cdot
g_{\cdot\cdot}}
\nonumber 
\\
\qq\qq\qq\qq
 \Delta \varphi =\,\frac{q}{\hbar}\, \oint A_\mu \dot{x}^\mu 
 &{\color{blue}\mbox{A-B}}\,|\,{\color{blue}\mbox{proper time}}&c\Delta \tau=\oint\sqrt{g_{\mu\nu}\dot x^\mu\dot x^\nu}
 \eee
We would like to use differential equations to compute the fields produced by {\it moving} sources. To this end we define charge- and current-densities $j$ and the energy-momentum tensor $T$. 

The question: ``Is the electromagnetic force field $(\vec E/c,\,\vec B)$ or the field of potentials $(V/c,\vec A)$ the fundamental field?'' has been source to a long controversy. Some suggested that the acceleration of a small, point-like test charge $q$ of mass $m$ is observable. This acceleration is given by the Lorentz force in terms of the electromagnetic force field $(\vec E/c,\,\vec B)$. The latter is conveniently encoded in the ``field strength'' $F_{\mu \nu}$, an anti-symmetric $4\times 4$ matrix and computed from the first partial derivatives of the 4-potential $(V/c,\vec A)$. Later it has been recognized that distances are not observables and {\it a fortiori} accelerations are not observables. Also quantum mechanics, in particular the famous Aharonov-Bohm effect (A-B), tells us that the potential is fundamental, the force field is an auxiliary, derived field.

We write the field equations, the Maxwell and the Einstein equations, as second order partial differential equations with the potentials as unknowns and the source densities given. The potentials outside a lonely, static and spherically symmetric charge- or mass-distribution are the first exact solutions that one computes. For Maxwell we obtain a pure electric potential $V=Q/(4\pi\,\epsilon_0r)$ where $r$ is the distance to the center of the charge distribution. For Einstein, we obtain a gravitational potential with leading term $-MG/r$ plus a small term suppressed by $2MG/(c^2 r)$ and therefore falling off like $1/r^2$  plus a harmonic oscillator term $-{\textstyle\frac{1}{6}}c^2\Lambda r^2$. The first correction to Newton's universal law is responsible for the perihelium advance of Mercury. The second correction, for positive cosmological constant $\Lambda $, induces a repulsive force which increases linearly with distance and which can explain the accelerated expansion of our universe.

Of course in the last two sentences, we have already used the law governing the motion of test particles, the Lorentz force for Maxwell and the ``geodesic equation'' for Einstein.  Both laws give the acceleration in terms of first partial derivatives of the potentials $A$ and $g$. The geodesic equation therefore identifies the ``connection'' $\Gamma $ as gravitational force while the ``metric'' $g$ encodes the gravitational potentials. Note the mismatch: the electromagnetic force field is the curvature $F$ of the ``gauge connection'' or ``gauge potential'' $A$, while the gravitational force is the connection $\Gamma $ of the metric $g$. The curvature $R$ of the connection $\Gamma $ describes the gradient of the gravitational force. Since this gradient is responsible for the tides of the sea, $R$ is sometimes called tidal force. 

Physically the geodesic equation expresses Newton's first axiom: force = mass $\times$ acceleration. Note that the mass of the test particle cancels. This is the equivalence principle visualized impressively
in Newton's tube. Note that thanks to this cancelation Einstein's test particle can be massless. (This is only true for Maxwell's test particle in the trivial case $q=0$.) Therefore general relativity predicts the bending of light by a massive object like the sun. (Warning: the geodesic equation is only valid for spinless test particles. Including the spin of the photon gives rise to more complicated phenomena: gravitational birefringence, studied in recent work by Christian Duval who left us in september 2018.)

General relativity has three axioms, the Einstein equation, the geodesic equation (the latter is in fact a consequence of the former) and the axiom of time. This axiom will be discussed in the next section together with its electromagnetic analogue, the Aharonov-Bohm effect. 

To close this section, we mention a few important features of Maxwell's and Einstein's theories.

\vspace{2mm}
\begin{center}
\begin{tabular}{r c l}
{
8 additive terms}
&
{\color{blue} size of $\dd_2$}&
$\sim\,10^5$ add. terms in $g$ \& $g^{-1}$

 \\[2mm]

Poincar\'e, gauge&{\color{blue}symmetries}
& {
diffeomorphisms}\\[2mm]

$\pa_\mu j^\mu=0$&{\color{blue}
(non-)conservation}
&
$D_\mu T^{\mu\nu}=0$
\\[2mm]

linear, 2nd order&{\color{blue}uniqueness of $\dd_2$}&2nd order 

\\[2mm]
 $\epsilon_0=8.854187817\cdot10^{-12}\,{\rm N^{-1}C^2m^{-2}}$&{\color{blue}\mbox{coupl.\  const.}}&$G=6.674208\cdot10^{-11}\,{\rm N\,kg^{-2}m^2}$
\nonumber\\ 
$\pm\, 0\,\% $\qq\qq\qq\qq\qq&&\qq\qq\qq\qq\qq $\pm\, 5\cdot10^{-5}$
\nonumber\\
&&$\Lambda =1.11\cdot 10^{-52}\,{\rm m^{-2}}\pm\, 2\, \%$
\nonumber
\end{tabular}
\end{center}
\vspace{2mm}

Maxwell's operator $\dd_2$ (written in inertial coordinates)  has 8 additive terms in $A$. These terms were deduced from experimental facts and charge conservation. Einstein's operator $\dd_2$ has some $10^5$ additive terms in $g$ and its inverse. These are hidden in the Ricci curvature $R_{\mu \nu }$ and the curvature scalar $R$ and were deduced from invariance under general coordinate transformations and energy-momentum (non-)conservation. 

Maxwell's theory (including the Aharonov-Bohm effect) is invariant
under Poincar\'e transformations and under gauge transformations. Einstein's theory is invariant under general coordinate transformations (or ``diffeomorphisms'').

Maxwell's equations imply charge conservation, $\pa_\mu j^\mu=0$. Einstein's equations imply conservation of the energy-momentum of matter if there is no gravitational field. Otherwise there can be exchange of energy and momentum between matter and the gravitational field; one speaks about ``covariant'' (matter) energy-momentum conservation, $D_\mu T^{\mu\nu}=0$. It should be noted that the linearity of Maxwell's equations implies that the electromagnetic field does not carry its own source = charge, while the non-linearity of Einstein's equations implies that a generic gravitational field does carry its own source = energy-momentum. (Warning: there is still no consensus on a definition of the energy density of a gravitational field.)

Maxwell's operator is the unique 1-parameter family of {\it linear} second order partial differential operators acting on the potentials $A$ that are compatible with charge conservation and invariant under Poincar\'e and gauge transformations. This single parameter is the coupling constant $\epsilon_0$.

Einstein's operator is the unique 2-parameter family of second order partial differential operators acting on the metric tensor $g$ and that are compatible with covariant (matter) energy-momentum conservation and invariant under general coordinate transformations.  The two parameters are the coupling constants $G$ and $\Lambda $.

\section{Chronometric approach}

The twentieth century has witnessed two revolutions concerning time, a technological one and a conceptual one.

Before the technological revolution, the official time keeping device was the rotating Earth with two draw-backs: its extended size and the inherent relative error of $10^{-8}$ due to the Earth not being rigid enough. The daily period of Earth was replaced by the period of light emitted by a cesium 133 atom. Atomic clocks are point-like enough for travelling and  allowed for time keeping with relative accuracy of $10^{-16}$ in 1980, see Figure 2, taken from  Soffel's book \cite{soffel}. Today the accuracy is reaching $10^{-18}$.

\begin{figure}[h]
\begin{center}
\includegraphics[width=12.6cm, height=8.9cm]{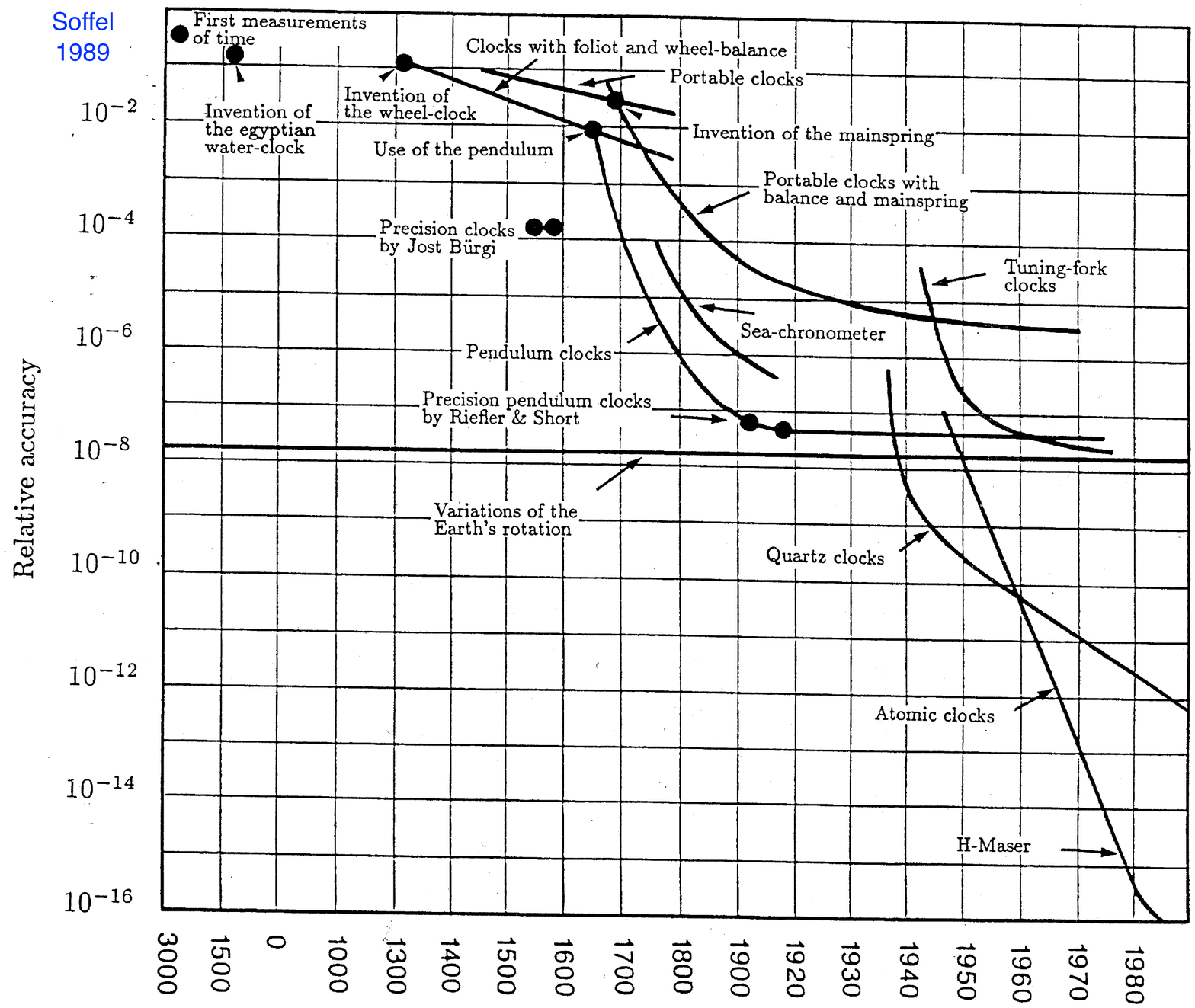}
\caption{The historical evolution of the accuracy in time keeping. Figure taken from Soffel's book \cite{soffel}. 
 }
\end{center}
\end{figure}

Before the conceptual revolution, time was absolute. Einstein taught us already in 1905 to be more careful. He defined proper time $\tau$ for a point-like clock travelling with velocity less than the speed of light on a trajectory $x^\mu (p)$ in a gravitational potential $g_{\mu \nu}(x)$:
\bb c\tau\dpp=\int_{p_0}^{p_1}\sqrt{
g_{\mu \nu}(x(p))\,\frac{\de x^\mu }{\de p}\,\frac{\de x^\nu }{\de p}}\ \de p.\ee 
Of course the signature of our metric is $+---$.

By definition proper time is proper to each clock and depends on its entire history: how much time it spent in what gravitational potential and with which velocities it was cruising. Today we know with very high precision that point-like clocks do indicate proper time. To test the axiom of time we need two precise clocks and synchronize them when they are at the same location. Then we make them travel on separate paths, reunite them and read their proper time difference $\Delta \tau$. It coincides with the line integral over the closed path defined by the two paths between synchronisation and reunification. 

The electromagnetic (quantum) analogue of proper time difference is the phase difference which the wave function of a charged particle accumulates when it passes through two holes in a screen and then follows two distinct paths in an electromagnetic potential. The two paths meet again and there the phase shift is measured. This phase shift can be computed from Schr\"odinger's equation coupled to the electromagnetic {\it potential} and is the line integral of this potential over the closed curve defined by the two paths between emission and reunification. The calculation was published  in 1949 by Ehrenberg  \& Siday \cite{es} and independently by Aharonov \& Bohm \cite{ab} in 1959. It was measured for the first time in 1960 by Chambers \cite{ch}.
 
The accuracy of length measurements evolved at a similar pace as the one of time measurements. The (French) definition of the meter as a part of the polar circumference of the Earth and materialized by a rigid rod of platinum was abandoned in favor of a multiple of the wave length of light emitted by a krypton 86 atom. However, as anticipated by Einstein, the precision of these new measurements killed the very notion of distance: rigid rods are incompatible with relativity as their speed of sound tends to infinity. But how useful can be a meter stick made of rubber?  Now the wavelength emitted by an atom is not rigid either and changes when the atom is immersed in an electric field as shown experimentally by Stark \cite{stark} in 1913. Einstein's axiom of time implies that the wavelength emitted by an atom also changes in a gravitational potential (explaining that Stark should have become an admirer of Einstein).

Both, rods and waves, are by definition extended objects and cannot be described by a single trajectory.

In 1983 the official funeral of the meter took place when  the  17th {\it Conf\'erence G\'en\'erale des Poids et Mesures} decided:
\begin{enumerate}
\item
 The metre is the length of the path travelled by light in vacuum during a time interval of 1/299\,792\,458 of a second.
\item
 The definition of the metre in force since 1960, based upon the transition between the levels 2p10 and 5d5 of the atom of krypton 86, is abrogated.
\end{enumerate}
At the same time, the speed of light became an absolute quantity {\it defined} by 299\,792\,458 meters per second and {\it without} error bars.

To see why this decision tolls the bell to distance measurements, let us take the distance between the Earth and the Moon. To simplify let us assume that they are both at rest. This distance is measured continuously since 1962 by the famous Lunar Laser Ranging experiment: A laser beam is sent to the Moon and reflected back to Earth. The time of the return flight, around 2.5 seconds, is measured and the distance Earth to Moon is obtained according to the definition in force: $\sim 3.8\,...\cdot 10^8$ m with a present error bar of $\pm \,1 $ cm. This distance is found to grow by 3.8 cm per year. If on the other hand we wanted to measure the distance Moon to Earth with both, laser cannon and clock, on the Moon (a more expensive set-up), we would obtain a distance increased by 30 cm due to the weaker gravitational potential at the Moon's surface, $d(E,m)\not=d(m,E)$.

\section{ Geometric approach}

Geometers like Gau\ss, Riemann, Christoffel, Levi-Civita, Ricci, Bianchi, ... worked out a definition of distance in curved spaces like the surface of the Earth. This definition, the arc-length of a geodesic, has the virtue of invariance under general coordinate transformations. This is precisely the virtue Einstein wanted for his theory of gravity. He simply replaced space by space-time, the arc length by proper time and kept the rest of the formalism. Therefore, although distances have been banned, the geometric language -- metric, connection, parallel transport, geodesics and curvature -- is still pertinent. The same language is pertinent to Maxwell's theory and to nonAbelian Yang-Mills theories describing the weak and strong nuclear forces. The next section tries to sketch out this common ground.

\section{Gauge theoretic approach}

To avoid subtleties of global geometry, let spacetime be a contractible open subset $M$ of $\rr^4$. Let $G$ be a finite dimensional Lie group. Define the gauge group $^MG$ as the set of all differentiable maps 
\bb
M &\longrightarrow &G \nonumber\\
x & \longmapsto & g(x) \nonumber
\ee
with pointwise multiplication $(g\tilde g)(x)\dpp=g(x)\tilde g(x)$.

The elements $g\in G$ are sometimes referred to as {\it rigid} transformations in contrast to the spacetime dependent gauge transformations $g(x)$. The gauge groupe $^MG$ is obtained by `gauging' or `mollifying' or `covariantizing' the rigid group $G$.

Both the gauge group $^MG$ and the diffeomorphism group $\mbox{\it Diff}(M)$, are infinite dimensional groups, not  Lie groups. Nevertheless they both admit infinite dimensional Lie algebras: $^M\gggg$ with  $\gggg$ the Lie algebra of $G$ and  commutators defined pointwise;  and the Lie algebra of vector fields with the Lie bracket.

 (Warning: some authors use the word gauge group in a wider sense including diffeomorphism groups and finite dimensional groups.)

\subsection{Linear algebra}

Fix a spacetime point $x\in M$ and consider its tangent space $T_xM=:V$. We have to parameterize the set of all Minkowski metrics  on $V$. Choose a basis $ b_i,\ i=0,\,1,\,2,\,3$ of $V$ and define the metric tensor by 
$$g_{ij}:=g(b_i,b_j).$$
{\bf Parameterisation 1:} For this fixed basis we have a one-to-one correspondence between the  set of all metrics and the set of all symmetric matrices of signature $(+---)$.\\[3mm]
The matrix ${g'}_{ij}$ of the metric $g$ with respect to a different
basis ${b'_i}$,
$$b'_i=(\gamma^{-1})^j_{\phantom{j}i}\,b_j,\qq \gamma \in \mbox{\it GL}_4,$$
is given by
$$g'_{ij}:=g(b'_i,b'_j)=(\gamma^{-1T}g\gamma^{-1})_{ij}.$$
Attention, we use $4\times 4$ matrices
to describe a metric as well as a change of bases, two quite different mathematical objects.\\[3mm]
{\bf Theorem} (Gram \& Schmidt):
Any metric has an orthonormal basis
$e_a$, i.e. a basis such that
\bb 
g(e_a,e_b)=\eta_{ab}:=
\begin{pmatrix}
+1&0&0&0\\
0&-1&0&0\\
0&0&-1&0\\
0&0&0&-1
\end{pmatrix}.
\ee
Given an orthonormal basis $e_a$, any other basis $e'_a$ with
\bb
e'_a=(\Lambda^{-1})^b_{\phantom{b}a}\,e_b,\quad\Lambda\in \mbox{\it GL}_4
\ee 
is also orthonormal if and only if
\bb
\eta=\Lambda^{-1T}\eta\,\Lambda^{-1},\qq \Lambda \in \mbox{\it O}(1,3)\subset \mbox{\it GL}_4.
\ee
{\bf Parameterisation 2:} Choose a basis $e_a$ of $V$ and declare it orthornomal. This defines a metric. However two bases connected by a Lorentz transformation $\Lambda $ define the same metric.
 
\subsection{Connection and curvature}

 Consider all tangent spaces together. They define a family of vector spaces indexed by the points $x$ of spacetime $M$. By definition a (pseudo-Riemannian) metric on $M$ is a differentiable family of metrics on each tangent space in the same sense as a vector field is a differentiable family of vectors. A frame is a differentiable family of bases $b_i(x)$ and an orthonormal frame $e_a(x)$ is ...
 
 Choose coordinates $x^\mu ,\ \mu=0,\,1,\,2,\,3$ on spacetime. They define a particular frame 
 \bb
  b_\mu \dpp =\,\frac{\pa}{\pa x^\mu}\,, 
  \ee
 called {holonomic}.\\[3mm]
 {\bf Theorem:} $M$ is flat if and only if it admits an orthonormal and holonomic frame. The corresponding coordinates are called intertial.\\[3mm]
{\bf Theorem:}
There is a unique metric and {\it torsion}-free connection $\Gamma $. It is a 1-form with values in the Lie algebra $\mbox{\it gl}_4$
 and can be computed explicitly in terms of $b_i(x)$, $g_{ij}(x)$  and their first partial derivatives with respect to an arbitrary coordinate system. With respect to a change of frames $\gamma (x)\in\,^M\mbox{\it GL}_4$ the connection transforms as
 \bb
  \Gamma '=\gamma \Gamma \gamma ^{-1}+\gamma \de \gamma ^{-1}.
  \ee
 {\bf Theorem:} The (Riemann) curvature 
 \bb
 R\dpp=\de\Gamma +{\textstyle\frac{1}{2}} [\Gamma ,\Gamma ]
\ee
is a 2-form with values in the Lie algebra $\mbox{\it gl}_4$ and transforms {\it homogeneously} under $\mbox{\it GL}_4$ gauge transformations:
 \bb
 R'=\gamma R\gamma ^{-1}.
 \ee
 
 \subsection{Three formulations of general relativity
}

For generic $M$ there are {\it three parameterizations} of the metric:
\begin{itemize}
\item
No name (no practical use): Choose an arbitrary frame $b_i(x)$ and a coordinate system $x^\mu$. Compute the metric tensor 
$g_{ij}$, the connection $\Gamma $ and the curvature $R$.  Changes of frames are $\mbox{\it GL}_4$ gauge transformations and changes of coordinates are diffeomorphisms. They decouple in the sense of the semi-direct product: $ ^M\mbox{\it GL}_4 \rtimes \mbox{\it Diff}(M)$.
\item
Einstein: Choose a coordinate system $x^\mu$ and use the metric tensor $g_{\mu \nu}(x)$. Now a change of coordinates induces a change of frames with the $\mbox{\it GL}_4$ gauge transformation given by the Jacobian of the diffeomorphism:
\bb
{\gamma^\mu }_\nu =\,\frac{\pa x^\mu }{\ \ \pa x^\nu}\,.
\ee 
``The $\mbox{\it GL}_4$ gauge group is lost due to gauge fixing''.
\item
Elie Cartan: Choose an orthonormal frame $e_a(x)$ and a coordinate system $x^\mu$. The metric tensor is constant, $\eta_{ab}$. The connection, traditionally denoted by $\omega $, now takes values in the Lorentz algebra $\mbox{\it so}(1,3)$ and is often called `spin connection'. 
\bb
\omega '=\Lambda \omega \Lambda ^{-1}+\Lambda \de \Lambda ^{-1}.
\ee
The symmetry group is $ ^M\mbox{\it O}(1,3) \rtimes \mbox{\it Diff}(M)$.
\end{itemize}
{\bf Theorem:} For a group $G$, denote by $G^{id}$ the connected component of the identity.\\
 ${\mbox{\it GL}_4}^{\it id}$ is doubly connected. Its covering group has no faithful finite dimensional represention.\\
$\mbox{\it SO}(1,3)^{\it id}$ is doubly connected. Its covering group $\mbox{\it Spin}(1,3)^{\it id}$ does have faithful finite dimensional representions.\\[3mm]
This theorem makes Cartan's parametrization mandatory if half-integer spin fields, e.g. Dirac particles, are present.\\[3mm]
{\bf Torsion} can be added simply in Cartan's formulation by declaring the spin connection $\omega$ to be an additional fundamental field, independent of the metric described by the orthonormal frame $e$. Then, by Cartan's equation, the half-integer spin densities are the source of torsion in the same sense that by Einstein's equation the energy-momentum densities are the source of curvature. And we have the commuting diagram shown in Figure 3. 

 \begin{figure}[h]
\begin{center}
\includegraphics[width=10.6cm, height=8.9cm]{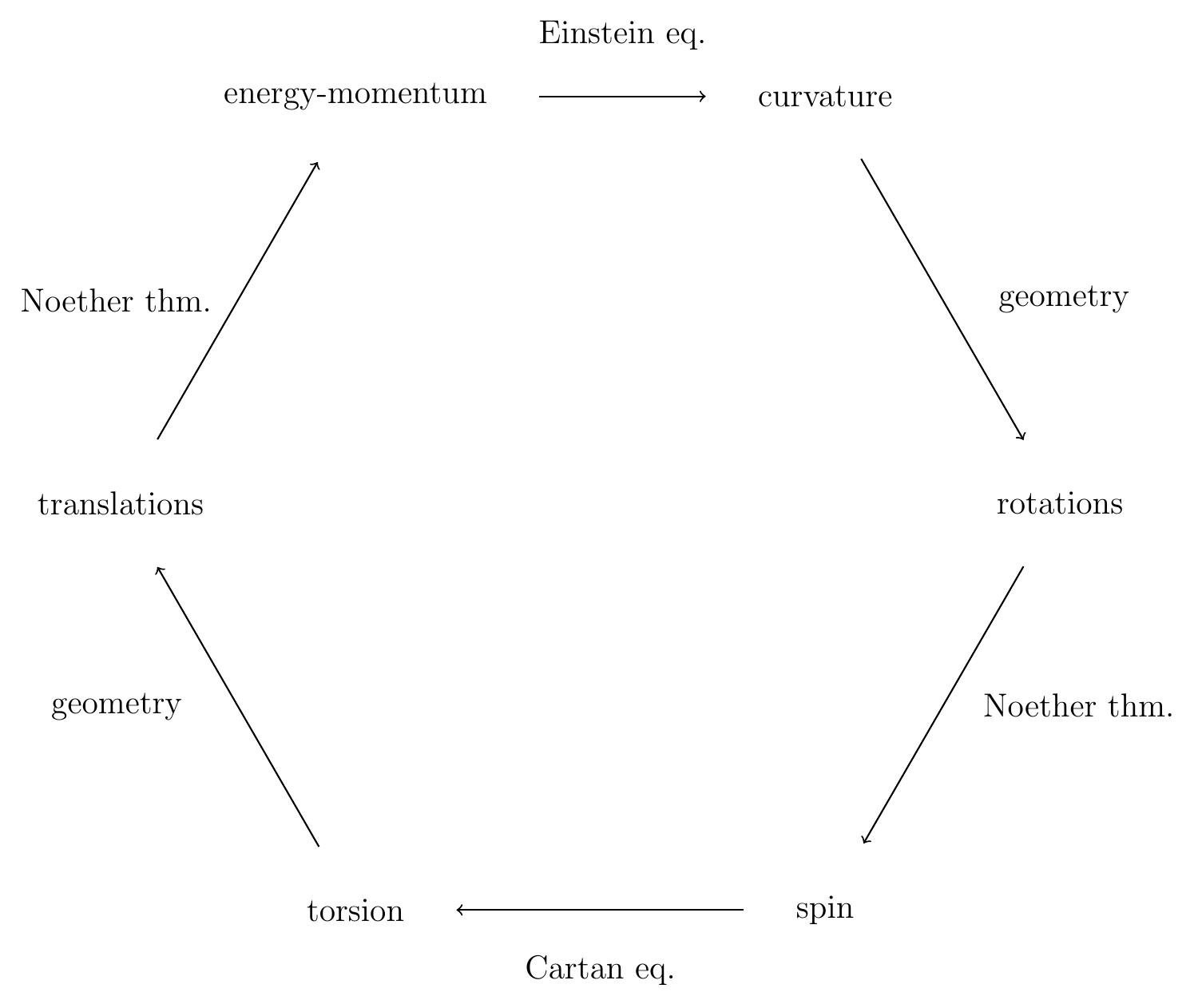}\caption{A commuting diagram }
\end{center}
\end{figure}

For details the reader is referred to \cite{gs}.

\section{Concluding remarks}

Today, experimental evidence forces us to abandon the concepts of rigidity and staticity on scales ranging from atomic to cosmic. This was already foreseen by Einstein's theories of relativity. (However he had a hard time to accept the observational evidence that our universe is not static.) In this context it is natural to also abandon rigid transformations and to favor gauge theories.

All current models describing the four basic forces are of Euler-Lagrange type and are gauge theories in the sense of the definition at the beginning of Section 5. The main differences between general relativity and the standard model of electromagnetic, weak and strong forces are:
\begin{itemize}\item
The fundamental field of general relativity (no torsion) is the gravitational potential = metric and the force is the connection = gauge potential; the fundamental field of the standard model is the gauge potential and the force is the curvature.
\item
The Lagrangian of general relativity is linear in curvature; the Lagrangian of the standard model is quadratic in curvature.
\item
The group $G=O(1,3)$ to be gauged in general relativity is simple but not compact;\\ the group $G=[SU(2)\times U(1)\times SU(3)]/[\zz_2\times \zz_3]$ to be gauged in the standard model is not simple but compact with finite dimensional unitary representations.
\item
In general relativity, masses are put in by hand; in the standard model, masses are generated by spontaneous symmetry breaking.
\item
The standard model is perturbatively renormalizable; general relativity -- to the best of our knowledge -- is not.
\end{itemize}
After a century of efforts, we are still lacking a unified model of all basic forces and we are still lacking a quantum theory of general relativity.
 

${}$\vspace{6mm}\\
\noindent
{\bf Acknowledgements:}
It is a pleasure to thank Silvia De Bianchi and Claus Kiefer for having organized a most stimulating conference and for their warm hospitality in Bad Honnef.
 
 This work has been carried out thanks to the support of the OCEVU Labex
(ANR-11-LABX-0060) and the A*MIDEX project (ANR-11-IDEX-0001-02) funded
by the "Investissements d'Avenir" French government program managed by
the ANR.

\end{document}